\documentclass{aa}
\usepackage{psfig, wrapfig}

\begin{document}



\title{GRB 040403: a faint X-ray rich  Gamma-ray Burst discovered
by INTEGRAL\thanks{Based on observations with INTEGRAL, an ESA
project with instruments and science data centre funded by ESA
member states (especially the PI countries: Denmark, France,
Germany, Italy, Switzerland, Spain), Czech Republic and Poland,
and with the participation of Russia and the USA. }}

  \author{S. Mereghetti\inst{1}, D. G\"{o}tz\inst{1}$^{,}$\inst{2},
  M.I. Andersen\inst{3},
  A. Castro-Tirado\inst{4},
  F. Frontera\inst{5,6},
  J. Gorosabel\inst{4},
  D. H. Hartmann\inst{7},
  J. Hjorth\inst{8},
  R. Hudec\inst{9},
  K. Hurley\inst{10},
  G. Pizzichini\inst{6},
  N. Produit\inst{11},
  A. Tarana\inst{12},
  M. Topinka\inst{9},
  P. Ubertini\inst{12},
  A. de Ugarte\inst{4}
     }

   \offprints{S. Mereghetti, email: sandro@mi.iasf.cnr.it}

  \institute{Istituto di Astrofisica Spaziale e Fisica Cosmica -- CNR,
                Sezione di Milano ``G.Occhialini'',
          Via Bassini 15, I-20133 Milano, Italy
          \and
              Dipartimento di Fisica, Universit\`{a} degli Studi di Milano Bicocca,
              P.zza della Scienza 3, I-20126 Milano, Italy
        \and
         Astrophysikalisches Institut Potsdam, An der Sternwarte 16,
         D-14482 Potsdam, Germany
        \and
          Instituto de Astrof\'{\i}sica de Andaluc\'{\i}a (IAA-CSIC),
         Apartado de Correos, 3004, 18080 Granada, Spain
       \and
       Physics Department, University of Ferrara, via Paradiso 12, I-44100 Ferrara, Italy
       \and
            Istituto di Astrofisica Spaziale e Fisica Cosmica -- CNR,
                Sezione di Bologna,
          Via Gobetti 101, I-40129 Bologna, Italy
        \and
                 Clemson University, Department of Physics \& Astronomy, Clemson, SC 29634-0978, USA
         \and
         Niels Bohr Institute, Astronomical Observatory, University of Copenhagen,
      Juliane Maries Vej 30, DK-2100 Copenhagen, Denmark
       \and
       Astronomical Institute,
       Academy of Sciences of the Czech  Republic,
       251 65 Ondrejov, Czech Republic
       \and
        University of California at Berkeley,
        Space Sciences Laboratories, Berkeley, CA 94720-7450
  \and
              Integral Science Data Centre, Chemin d'\'{E}cogia 16, CH-1290 Versoix, Switzerland
         \and
            Istituto di Astrofisica Spaziale e Fisica Cosmica -- CNR, Sezione di Roma, Via Fosso del Cavaliere 100, I-00133 Roma, Italy
   }


\abstract{GRB 040403 is one of the faintest gamma-ray bursts for
which a rapid and accurate localization has been obtained.  Here
we report on the gamma-ray properties of this burst, based on observations
with the IBIS instrument aboard INTEGRAL, and the results of searches for
its optical afterglow. The steep spectrum (power law photon index =
1.9 in the 20-200 keV range) implies that GRB 040403 is most
likely an X-ray rich burst. Our optical limit of R$>$24.2 at 16.5
hours after the burst, indicates a rather faint afterglow, similar to those
seen in other relatively soft and faint bursts.
  }

\authorrunning{S. Mereghetti  et al.}
\titlerunning{GRB 040403}

\maketitle

%

\section{Introduction}

Gamma-ray bursts (GRBs) are extremely energetic cosmic explosions
associated with the death of massive stars (see, e.g.,
\cite{review} for a recent review). Multi-waveband observations,
made possible by accurate (arc minutes) and fast (minutes)
localizations of the rapidly fading gamma-ray emission, are
crucial for the development of an understanding of this
phenomenon. Rapid follow-up observations also provide a unique
opportunity to probe the early universe, as typical GRB redshifts
are of order unity. Currently, fast localizations are obtained for
a few dozen GRBs per year, based on detections with HETE II and
INTEGRAL (\cite{hete}; \cite{ibasres}). A significant increase in
this rate is expected with the imminent launch of the Swift
satellite (\cite{swift}). It has been pointed out that, thanks to
the high sensitivity of its IBIS instrument (\cite{ubertini}), the
sample of INTEGRAL GRBs is likely to contain a large fraction of
bursts at high redshift (\cite{gorosabel}). In this respect, the
faintest and spectrally softest bursts are the most promising
cosmological probes.

On 2004 April 3, a faint GRB triggered the INTEGRAL Burst Alert
System (IBAS, \cite{ibas}). An IBAS Alert Packet containing the
burst coordinates with an uncertainty of only 2.8$'$ was
automatically distributed at 05:08:29 UT. The first photons from
the burst had reached the INTEGRAL satellite only 30 s earlier.
This is therefore one of the most precise and rapid GRB
localization obtained to date. A quick look analysis of the data
confirmed the automatically derived position and indicated some
evidence for a relatively soft spectrum (\cite{040403D}). A
refined analysis, announced within three hours of the GRB event,
reduced the positional uncertainty to 2.1$'$. Despite being at a
Galactic latitude (b=30$^{\circ}$) higher than the majority of
INTEGRAL bursts (\cite{ibasres}) and thus barely affected by
interstellar extinction (A$_{V}\sim$0.3), optical follow-ups of
GRB 040403 were somewhat discouraged by the presence of a full
moon.

Here we present a detailed analysis of the INTEGRAL data and the
results of searches for the optical afterglow carried out at the
Observatory of Sierra Nevada (OSN) and at the Nordic Optical
Telescope (NOT). We also report on pre-GRB images obtained with
the Burst Alert Robotic Telescope (BART) at the Ondrejov
Observatory, which allow us to place constraints on the presence
of optically variable sources before the onset of the prompt GRB
emission. To our knowledge, GRB 040403 was not detected by other
satellites,\footnote{two other bursts detected on the same day by
other satellites have been erroneously named in \cite{gcn2566}:
they occurred after the INTEGRAL burst reported here, and should
therefore be named GRB040403B and GRB040403C.} and no follow up
observations were performed in the X--ray and radio ranges.

\section{INTEGRAL observations}

The data presented here were obtained with ISGRI (\cite{lebrun}),
the lower energy detector of the IBIS imaging instrument. Due to
its 15 keV - 1 MeV  energy range, large field of view, high time
resolution and good spatial resolution, ISGRI provides the best
data used by IBAS for real time search and localization of GRBs.

\begin{figure}[t]
\psfig{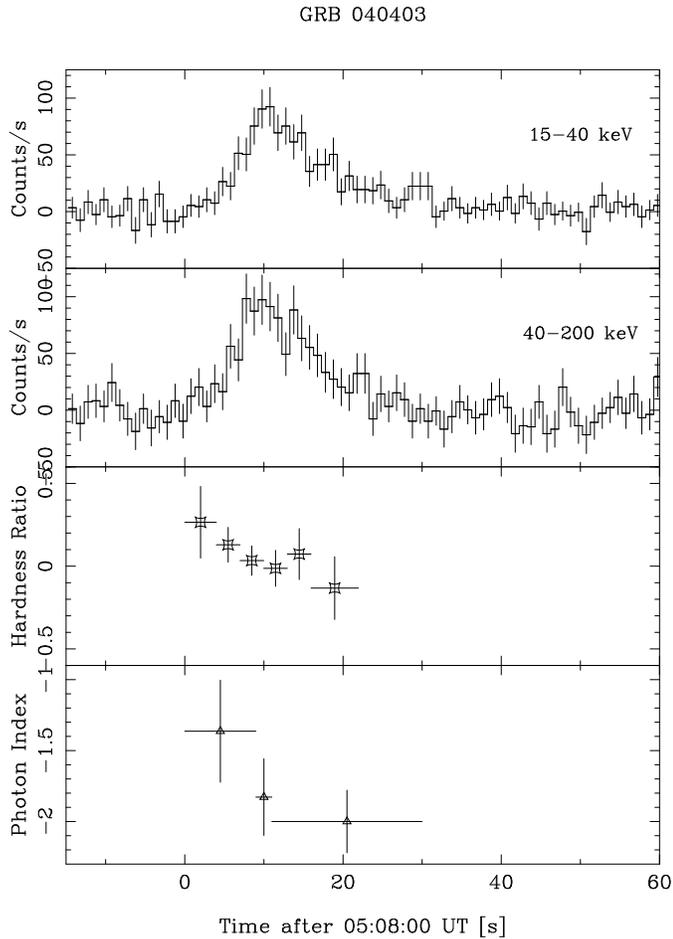} \caption{Light curves and
spectral evolution of GRB 040403:  (a) light curve obtained with
IBIS/ISGRI in the 15-40 keV energy range, binned in intervals of 1s
(only detector elements illuminated by more than 50\% by the source
were used); (b) same as (a) for the 40-200 keV range; (c) ratio of
15-40 keV to 40-200 keV counts; and (d) the power law photon index
obtained from the spectral analysis of three distinct time intervals}
\label{lc}
\end{figure}

GRB 040403 occurred at off-axis angles of Z=--4.28$^{\circ}$ and
Y=--4.34$^{\circ}$, in the fully coded region of the instrument
field of view\footnote{IBIS is a coded mask telescope. Sources
outside the fully coded field of view project an
aperture-modulated flux only on a fraction of the detection
plane.}. The top panels of Fig.\ref{lc} show the light curves in
the soft (15-40 keV) and hard (40-200 keV) ranges. The burst
profile exhibits a single peak with a rise time of about 5 s and a
slower decay. The T$_{90}$ duration is 19 s. The decay time scale
is longer in the softer energy band, indicating the presence of
spectral hard-to-soft evolution, as observed in many GRBs. This is
demonstrated by the hardness ratio evolution (third panel of
Fig.\ref{lc}) and by the time resolved spectral analysis described
below. We computed the cross correlation between the light curves
at energies below and above 40 keV, and found  that the soft light
curve has a time lag of 0.6$\pm$0.1 s.

\begin{figure}[t]
\psfig{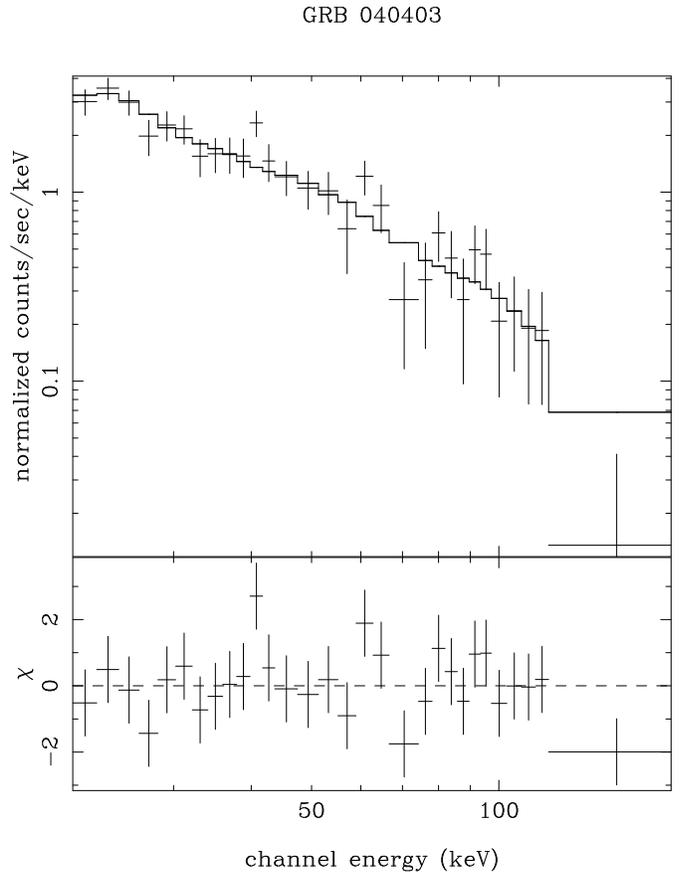}

\caption{IBIS/ISGRI spectrum of  GRB 040403 fitted with a power
law. Upper panel: data and best fit model. Lower panel: residuals
from the best fit spectrum in units of sigma.
 }

\label{sp}
\end{figure}

In order to measure the time integrated spectrum and the fluence
we extracted the counts in a 30 s long time interval starting at
5:08:00 UT. This yielded about 2500 net counts. For the spectral
analysis we used the most recent response matrix, which takes into
account the effects of the off-axis angle dependence of the mask
transparency. The spectrum was rebinned in order to have at least
20 counts per bin and fitted over the 20-200 keV range. A good fit
was obtained with a power law of photon index $\Gamma=1.90\pm0.15$
and 20-200 keV flux of 0.2 photons cm$^{-2}$ s$^{-1}$
($\chi^{2}$/dof = 12.39/12, see Fig.\ref{sp}). The fluence in the
same energy range was 5.0$\times10^{-7}$ erg cm$^{-2}$ and the
peak flux $\sim$0.50 photons cm$^{-2}$ s$^{-1}$ (over a 1s time
interval). Time resolved spectral analysis yields the power law
photon indexes shown in the lowest panel of Fig.\ref{lc},
confirming the hard-to-soft spectral evolution derived from the
hardness ratio.

An acceptable fit to the time averaged spectrum could also be
obtained with a Band function (\cite{band}), which has two more
free parameters than the simple power law adopted above.  However,
owing to the limited statistics and reduced energy range the fit
parameters cannot be constrained. The best fit is found for
$\alpha\sim$ --1.5, E$_p$$\sim$60 keV, and $\beta>$--3.

We also recomputed the GRB position by producing images in
different energy ranges and time intervals. The highest signal to
noise ratio was obtained in the 15-100 keV range and the
corresponding position of GRB 040403 is $\alpha_{J2000}$ = 7$^h$
40$^m$ 54$^s$, $\delta_{J2000}$ = +68$^{\circ}$ 12$'$ 55$''$. This
position, based on the final attitude reconstruction derived for
the satellite, is consistent with and supersedes the refined
position reported less than three hours after the GRB event
(\cite{040403D}). The 90\% confidence level error radius for the
GRB position is 2$'$. GRB 040403 was not detected above 120 keV,
consistent with the soft spectrum discussed above.

\begin{figure}[t]
\psfig{figure=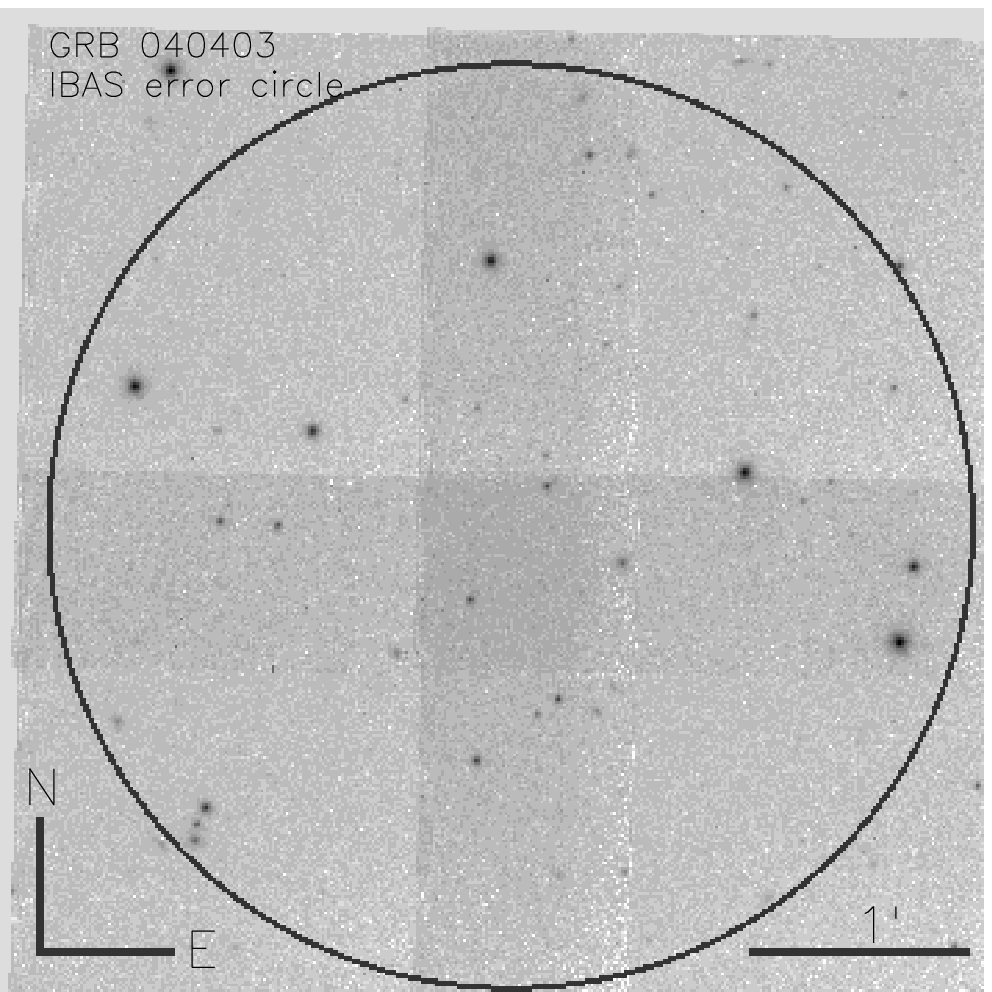,width=8.5cm}

\caption{The R-band image of the GRB~040403 error box, obtained
with the Nordic Optical Telescope (NOT) on April 4, 17 hours after
the GRB. The 2$\times$2 mosaic image covers the entire 2.1$'$ IBAS
error circle. The 3$\sigma$ limiting magnitude for this field is
R$_{\rm lim}$ = 24.2.}

\label{not}
\end{figure}

\section{Optical observations}

\subsection{Pre-burst imaging}

The robotic telescope BART (\cite{bart}) observed the position of
GRB 040403 on April 3, 2004 with two wide-field cameras. Images of
the GRB position were obtained at 03:17:32 and 03:36:01 UT
(mid-points of 120 s long exposures) i.e. 112 and 93 min before
the burst. No variable optical object was detected within the GRB
error box. Relatively poor weather conditions limit our ability to
reach very faint magnitudes. The magnitude constraints on any
pre-GRB emission obtained for these images are R$>$12 and
I$>$11.6. The second image was the last usable observation before
images started to get overexposed due to the rising sun.

\subsection{Search for optical afterglow emission}

Following the IBAS trigger, we observed the field of GRB 040403 at
optical wavelengths starting April 3.83 UT, with the 1.5 m
telescope at the OSN in Granada and with the 2.5 m NOT telescope
in La Palma. The OSN images in the R band (6 $\times$ 600 s) were
obtained with the CCD VersArray 2048B which provides a 7\arcmin
$\times$ 7\arcmin~ field of view. Comparison images were acquired
on 4 and 20 April 2004 (9 $\times$ 600 s).

The NOT observations were obtained with the Standby camera
(StanCam). As the field of view of StanCam is only 3 arcmin
square, a 2$\times$2 mosaic was acquired, in order to cover the
entire error-box, each mosaic element consisting of three
integrations of 300 s. The observations were initiated on April
3.890 UT, with a mean epoch of April 3.940 UT. The seeing was
$\sim$0.80$''$ resulting in a 3$\sigma$ limiting magnitude of
R=24.2. During the following night, second epoch imaging was
carried out, but due to poor seeing and partial cloud cover, it
was not nearly as deep as the images from the previous epoch.
Visual comparison did not reveal any variable object. Subsequent
NOT observations, during a possible SN re-brightening phase 10
days after the GRB, were planned but not conducted, due to
enduring bad weather. To fully exploit the depth of the first
epoch observations, a deeper R-band image was acquired on June 21,
2004, using ALFOSC at the NOT, which covers the entire error box
in a single exposure. The resulting image, which was composed of
five 300~s exposures, obtained under seeing of 0.85$''$, reaches
half a magnitude deeper than the first epoch images. Visual
comparison of the residual image after image subtraction does not
reveal any variable source to the detection limit of the first
epoch. The position dependent variable PSF of the ALFOSC images
effectively prevents quantitative image subtraction from being
employed on this data set. We decided not to combine the OSN and
the NOT data, as the difference in depth and seeing is too large
to allow a significantly lower limit from combined data. Several
galaxies fainter than R$\sim$20 contained in the INTEGRAL error
region could be considered as the potential GRB host, but, lacking
a more precise position, none can be singled out.

\section{Discussion}

GRB 040403 is one of the faintest GRBs detected by INTEGRAL to
date. The soft spectrum measured with ISGRI indicates that it
likely belongs to the class of X-ray rich GRBs. The distinction
between normal GRBs, X-ray rich GRBs and X-ray Flashes (XRF) is
somewhat arbitrary, but a widely used definition is based on the
ratio of the fluences in the 2-30 and 30-400 keV energy ranges, S
= Log(F$_{2-30}$/F$_{30-400}$). XRFs have S$>$0, normal GRBs have
S$<$--0.5, and bursts with intermediate values are considered
X-ray rich (\cite{hete}). We do not have information on the GRB
040403 spectrum below 15 keV\footnote{The burst was outside the
field of view of the JEM-X instrument (4-35 keV, \cite{lund}) and
of the optical camera (OMC, \cite{mashesse}) aboard INTEGRAL} and
we have only ISGRI upper limits above $\sim$150 keV. Assuming that
the GRB spectrum is well represented by a single power law over
the entire relevant range, we infer S = --0.11$\pm$0.18. Only a
significant spectral flattening to a photon index of $\sim$0.5
below $\sim$20-30 keV would result in values of S closer to those
observed in typical GRBs. Also adopting Band spectra with various
parameter values in the range allowed by our spectral fits, we
obtain values of S larger than --0.5. We thus conclude that GRB
0040403 was most likely a member of the class of X-ray rich GRBs.

Our afterglow limit of R$>$24.2 at 16.5 hours after the burst and
those reported soon after the event (R $\geq$ 19 at 1.1 hr;
\cite{moran}) are rather constraining, compared to the majority of
limits obtained to date for other bursts. Most of the observed GRB
afterglows are brighter than R$\sim$23 at t=1 day (\cite{fynbo};
\cite{fox}; \cite{berger}). It is interesting to note that a
similarly faint optical afterglow was reported also for GRB
030227, another faint and spectrally soft burst discovered by
INTEGRAL (\cite{030227P}; \cite{030227O}).

High redshift could be responsible for the faintness of the
optical afterglows and the X-ray richness of these faint INTEGRAL
bursts. Based on the correlation between spectral lag and
luminosity  established for GRBs (\cite{norris}), the 0.6 s lag
measured in GRB 040403 corresponds to an isotropic luminosity of
$\sim$1.2$\times$10$^{51}$ erg s$^{-1}$. For such a luminosity,
the flux measured by INTEGRAL implies a redshift z=2.1\footnote{we
have assumed H$_{o}$=65 km s$^{-1}$ Mpc$^{-1}$,
$\Omega_{\Lambda}$=0.7 and $\Omega_{m}$=0.3}. More extensive
prompt multi-wavelength follow-up observations of the IBAS Alerts
for faint bursts are clearly needed to assess whether INTEGRAL is
indeed preferentially sampling the farthest GRBs.

We finally note that the upper limit from the image obtained on
April 20 also excludes the presence of a nearby (z = 0.1-0.2)
underlying, non-obscured supernova similar to SN 1998bw
(\cite{galama}) in the error box. It is now fairly well
established that most long-duration GRBs are followed by extra
light, commonly attributed to an associated supernova
(\cite{ZKH}). This link is a natural consequence of the collapsar
model (\cite{woosley}), and GRB 030329 is so far the best direct
proof (\cite{stanek}; \cite{hjorth}; \cite{Greiner}) of this
supernova-GRB association. HETE-2 has provided strong evidence
that X-ray flashes (XRFs), X-ray rich GRBs, and GRBs are in fact
drawn from the same underlying source population (\cite{LDG}). It
is thus natural to assume that X-ray rich GRBs are also followed
by emission from a supernova. The lack of deep late time
observations for this GRB unfortunately does not allow us to test
this hypothesis, but we are very optimistic that future rapid GRB
localizations with INTEGRAL, HETE-2, or Swift will eventually
reveal enough   to decide whether or not the "XRF-GRB continuum"
hypothesis holds.

\begin{acknowledgements}
This research has been partially supported by the Italian Space
Agency. K.H. and D.H. are grateful for support under NASA Grant
NAG5-12706 in the INTEGRAL US Guest Investigator program.
\end{acknowledgements}

\end{document}